\documentclass[useAMS,usenatbib,letterpaper]{mn2e}
\usepackage{times}
\usepackage{epsfig,graphicx,graphics}

\title[Hyperfine splitting of Al\,{\sc vi}~3.66um]{Hyperfine splitting
 of [Al\,{\sc vi}]~3.66$\mu$m and the  Al isotopic ratio in NGC~6302} 

\author[S. Casassus et al.]{S. Casassus$^{1}$\thanks{E-mail:
simon@das.uchile.cl (SC)}, P. J. Storey$^{2}$, M. J. Barlow$^{2}$,
P. F. Roche$^{3}$\\ $^{1}$ Departamento de Astronom\'{\i}a,
Universidad de Chile, Casilla 36-D, Santiago, Chile\\ $^{2}$
Department of Physics and Astronomy, University College London, Gower
Street, London WC1E 6BT\\ $^{3}$ Denys Wilkinson building, Physics
Department, Oxford University, Keble Road, Oxford OX1 3RH}

\begin{document}

\date{2004 August 20}

\pagerange{\pageref{firstpage}--\pageref{lastpage}} \pubyear{2002}

\maketitle

\label{firstpage}

\begin{abstract}
The core of planetary nebula NGC 6302 is filled with high-excitation
photoionised gas at low expansion velocities. It represents a unique
astrophysical situation in which to search for hyperfine structure
(HFS) in coronal emission lines from highly ionised species.  HFS is
otherwise blended by thermal or velocity broadening.  Spectra
containing [Al\,{\sc vi}]~3.66$\mu$m $^3$P$_2 \leftarrow ^3$P$_1$,
obtained with Phoenix on Gemini-South at resolving powers of up to
75000, resolve the line into five hyperfine components separated by 20
to 60~km~s$^{-1}$ due to the coupling of the $I=5/2$ nuclear spin of
$^{27}$Al with the total electronic angular momentum $J$.  $^{26}$Al
has a different nuclear spin of $I=5$, and a different HFS, which
allows us to place a 3~$\sigma$ upper limit on the $^{26}$Al/$^{27}$Al
abundance ratio of 1/33.  We measure the HFS magnetic-dipole coupling
constants for [Al\,{\sc vi}], and provide the first estimates of the
electric-quadrupole HFS coupling constants obtained through
astronomical observations of an atomic transition.

\end{abstract}

\begin{keywords}
atomic data -- atomic processes -- line:identification -- line:
profiles -- ISM: abundances -- platenary nebulae: NGC~6302.
\end{keywords}

\section{Introduction}

%hyperfine levels. While molecular hyperfine transitions are common in
%the radio range, at shorter wavatomic hyperfine structure (HFS) has
%seldom been observed other than in solar absorption lines from neutral
%species \citep{abt52}, such as Mn\,{\sc i}, whose transitions at

The interaction between the electronic wave-function and a non-zero
nuclear magnetic dipole splits a fine-structure level $\{L,J\}$ into
hyperfine levels. While hyperfine transitions are common in the radio
range, at shorter wavelengths atomic hyperfine structure (HFS) has
seldom been resolved in emission.  Examples of hyperfine broadening
include solar absorption lines from neutral species \citep{abt52},
such as Mn\,{\sc i}, whose transitions at 1.7743~$\mu$m are a rare
example of unblended hyperfine lines \citep{mel99}. \citet[and
references therein,][]{boo83} summarise the effects of HFS on stellar
spectra: its neglect results in incorrect measures of line broadening,
and mismatched curves of growths, leading to $\sim$0.2~dex errors in
the inferred photosphere elemental abundances, even for faint lines
far from saturation.  Both HFS and saturation lead to line
broadening. This degeneracy complicates the use of stellar absorption
lines as diagnostic of the hyperfine coupling constants.  Hyperfine
splitting can also be resolved in interstellar Na\,{\sc i} D-line
absorption in the local {\sc ism}, which requires resolving powers of
$\ga 5~10^{5}$ \citep[][for more recent data]{way78,bar95}.

The 1.8~MeV gamma-ray emission due to the decay of $^{26}$Al into
$^{26}$Mg has been the object of extensive space borne surveys: with a
half-life of $7.2~10^{5}$~yr, $^{26}$Al is a signpost of recent
nucleosynthesis. Line emission at 1.8~MeV from the diffuse {\sc ism}
\citep[as observed by {\em COMPTEL},][]{die95} is consistent with an
$^{26}$Al source in either AGB stars \citep{for91}, novae, supernovae,
Wolf-Rayet stars \citep{pra04} or from cosmic-ray collisions in
molecular clouds \citep{cla94}.  The {\em INTEGRAL} \citep{win03}
mission holds the promise of improved angular resolution with which to
identify the most important contributor to the diffuse emission.

Although the decay of $^{26}$Al is observed in the ISM at large, the
$^{26}$Al/$^{27}$Al isotopic abundance ratio (hereafter
$R_\mathrm{iso}$) has never been measured in any astrophysical
source. The only available upper limit in any specific object is that
of \citet{ban04}, who observed the vibronic bands of AlO at 1.5~$\mu$m
in the nova-like variable V4332 Sgr, and reported an upper limit of
$\sim$1/10, lacking a statistical discussion.

As an application of our detection of HFS in [Al\,{\sc vi}]~3.66$\mu$m
$^3$P$_2 \leftarrow ^3$P$_1$ (hereafter [Al\,{\sc vi}]), which is the
first in an astrophysical near-IR emission line, we can set an upper
limit on $^{26}$Al/$^{27}$Al using the difference in the HFS of both
isotopes: the stable isotope $^{27}$Al has a nuclear spin $I=5/2$,
while $^{26}$Al has a nuclear spin $I=5$. This new upper limit is the
most stringent obtained so far in any astrophysical target.

The first detection of atomic HFS in emission, aside from the 21~cm
H\,{\sc i} line, is to our knowledge the observation of resolved HFS
in [$^{13}$C\,{\sc ii}] 157.8$\mu$m
$^{2}$P$_{1/2}\leftarrow^{2}$P$_{3/2}$ by \citet{bor88}. \citet{kel95}
identified multiple components in [Na\,{\sc iv}]~9.0$\mu$m with the
hyperfine splitting of $^{3}$P$_2\leftarrow ^{3}$P$_1$. Although
[Na\,{\sc iv}]~9.04$\mu$m and [Al\,{\sc vi}]~3.66$\mu$m are the same
fine-structure transitions from isoelectronic ions, they differ in
nuclear spin and electronic wave-functions, leading to different
hyperfine structures.

We also derive values for the electric quadrupole constants (hereafter
$B$ constants) in the [Al\,{\sc vi}] transition. To our knowledge,
this is the first measurement of such constants in an atomic
transition in any astrophysical object, although the $B$ quadrupole
constants have been measured in molecular transitions. In contrast
with atomic HFS, in molecules the hyperfine splitting of a given
rotational transition primarily derives from nuclear quadrupole
moments rather than from nuclear magnetic moments
\citep[e.g.,][]{tow55}.  For instance $B$ values have previously been
measured by \citet{tur75} in CN(K = 1 -0) at 2.6~mm, and by
\citet{ziu92} in HCNH$^+$(J = 1-0) at 74~GHz. In this work we show
that the inclusion of the atomic electric quadrupole terms has
important spectroscopic consequences.  It allows improved measurement
of the magnetic dipole coupling constants by lifting the statistical
bias between the magnetic dipole constants of the upper and lower
levels.

In this work we demonstrate the use of HFS itself as a diagnostic tool
in the context of planetary nebulae (PNe).  Atomic hyperfine effects
have previously been used by \citet{cle97} in C\,{\sc
iii}]~$\lambda$1909.6 $^1$S\,$_0\leftarrow\,^3$P$_0$ to measure
$^{13}$C/$^{12}$C in PNe. They recognised that the non-zero nuclear
spin of $^{13}$C additionally\footnote{This multiplet arrises from the
mixing of $^3$P and $^1$P states due to magnetic interactions between
the electrons} mixes the $^3$P$_0$ and $^3$P$_1$ fine-structure
states.  C\,{\sc iii}]~$\lambda$1909.6 is dipolar-electric in
$^{13}$C, while it is completely forbidden in $^{12}$C because it has
no net nuclear spin. This C\,{\sc iii} multiplet is thus composed of
three lines, one of which is due solely to $^{13}$C.

%A 3000--10\,000~\AA~ echellogram we acquired with UVES on the VLT does
%not confirm the report of \citet{mea80} for broad wings under
%[Ne\,{\sc v}]~3426~\AA (Casassus et al., in preparation), which has
%been taken as evidence for a fast wind in NGC~6302.

NGC~6302 is the highest excitation PN known, with a spectrum rich in
molecular lines, dust, and coronal ions such as [Si\,{\sc
ix}]~3.93~$\mu$m, which can only be produced by photons harder than
303~eV, or by electron collisions at $T_e \approx 10^{6}~$K. Its
spectrum can be reproduced by ionisation-bounded photoionisation
models with a $T=250\,000~$K central star \citep{cas00}, and the
absence of a fast wind makes improbable a significant contribution
from shock excitation. Although the report of \citet{mea80} for broad
wings under [Ne\,{\sc v}]~3426~\AA has been taken as evidence for a
fast wind in NGC~6302, a 3000--10\,000~\AA~ echellogramme we acquired
with UVES on the VLT (Casassus et al., in preparation) does not
confirm the observations of \citet{mea80}. The photoionised coronal
lines in NGC~6302 are astonishingly narrow \citep{ash88} compared to
conditions of collisional ionisation where their abundance is maximum.
The line-widths measured by \citet{cas00} reflect negligible thermal
broadening from photo-ionised gas temperatures of 20\,000~K, and very
small expansion velocities in a filled-in nebula.

It is its small expansion velocity and rich spectrum that makes
NGC~6302 an ideal object for the use of hyperfine structure as a
diagnostic tool.

% our method to fit the hyperfine splitting and obtain a measure of the
%coupling constants

%We describe data acquisition (Section~\ref{sec:obs}), then data
%analysis and results (Section~\ref{sec:hfsfits}), and summarise our
%conclusions (Section~\ref{sec:conc}). Data reduction and analysis were
%carried-out using the Perl Data Language ({\tt http://pdl.perl.org}).

We describe data acquisition in Section~\ref{sec:obs}, then data
analysis and results in Section~\ref{sec:hfsfits}, and summarise our
conclusions in Section~\ref{sec:conc}. Data reduction and analysis
were carried-out using the Perl Data Language ({\tt
http://pdl.perl.org}).

\section{Observations} \label{sec:obs}

\begin{figure}
\includegraphics[width=8.5cm,height=!]{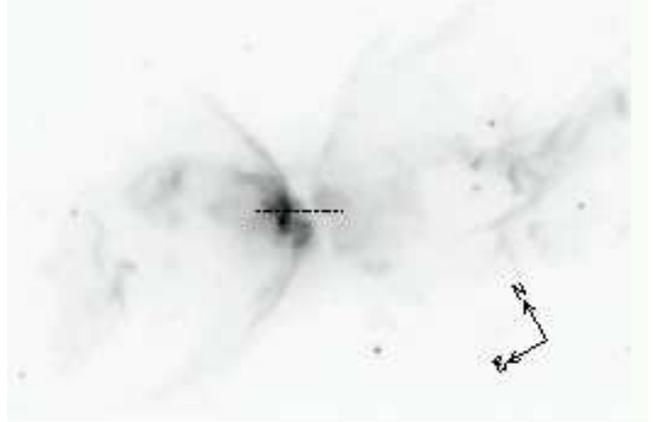}
 \caption{\label{fig:slit} Overlay of the Phoenix 14~$''$ slit on the R-band
 Gemini acquisition image.}
\end{figure}

We observed NGC~6302 with Phoenix \citep{hin03} on Gemini South on 5
nights of May and July 2003, as summarised in
Table~\ref{table:log}. The slit position angle was 70~deg East of
North, and it was centred on NGC~6302's radio core at J2000
RA=17:13:44.4, DEC=-37:06:11.2, as inferred from the 5~GHz map of
\citet{gom89}, at the position of the intensity decrease in the centre
of the putative radio torus. Fig.~\ref{fig:slit} shows the slit
position overlaid on the R-band image obtained with Gemini's
acquisition camera.  Background cancellation was obtained by
differentiation with a reference field devoid of nebular emission,
offset 40$''$ North of the nebular core.  Typical integration times in
the \mbox{[Al\,{\sc vi}]} settings were 1h-2h on-source for each
night, but the noise level largely reflects the weather conditions.
Poor weather also results in inaccurate background cancellation. The
seeing has a direct impact on the resolution of the spectra, by
convolving the emission in the slit with neighbouring emission from
the expanding nebula. The emission that falls through the slit is the
convolution of the slit aperture with the point spread
function. Because of the spatial variations of radial velocity within
the nebula, poor seeing allows emission from material with a wider
range of velocities to be admitted by the spectrograph slit.  The
resulting spectra are therefore degraded by a combination of the
instrumental resolution and the spatial variations in velocity.

%differential instrumental refraction.

%Atmospheric diffraction between the wavelength used for guiding and
%the observing wavelength is accounted for in the telescope control
%system.

%\footnote{peaking up means shifting the
%slit slightly to find the position of maximum signal}

The acquisition of a precise position in the nebula is important to
obtain consistent spectra. To centre the slit on the position of NGC
6302's radio core we peaked-up on a reference astrometric standard in
the K band, then offset to the object, switched detector settings and
applied an additional offset to account for the difference in
refraction between the filters.  Peaking-up with a narrow slit is
difficult because of variations in the seeing on time-scales
comparable to the acquisition procedure. The accuracy involved in
peaking-up depends on the seeing and the slit-width. The overall
positional uncertainty is $\sim$0.6~arcsec, as estimated by adding in
quadrature the errors involved in peaking up, of about 0.35$''$ (or
twice the slit-width), in offsetting from the reference star, of
$\sim$0.5~$''$, and in the filter change, of $\sim$0.1$''$ (Bernadette
Rodgers, private communication).

%# File ngc6302.fits is an Acquisition camera R-band image of the observed 
%# field.  Pixel location x=571. y=580. is the approximate position of the 
%# center of the phoenix slit (on-source position).  AcqCam pixel scale 
%# is 0.12 arcsec/pix and the orientation is N right, E down (but rotated
%# 20degrees (PA=70) such that N~4:00.  The slit runs vertically in the image.

\begin{figure}
\includegraphics[width=8.5cm,height=!]{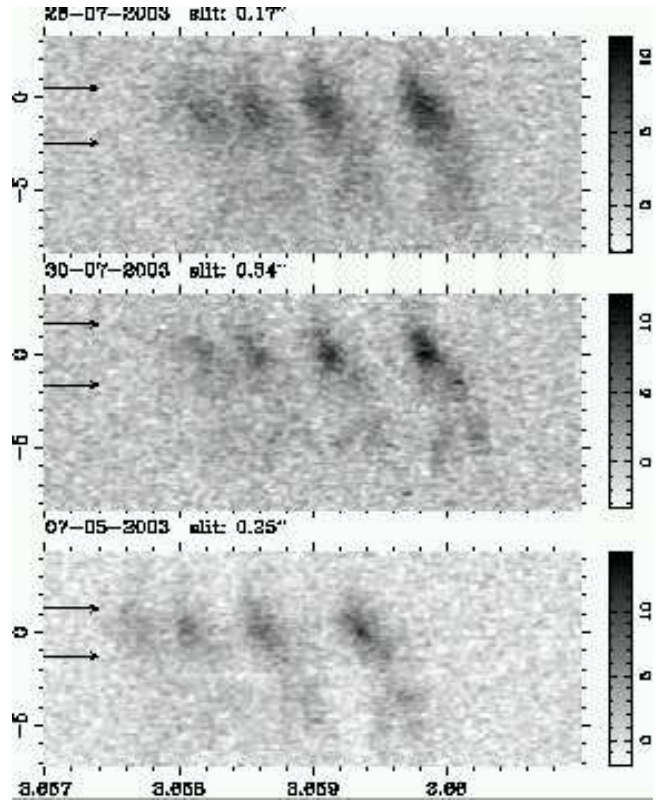}
 \caption{\label{fig:frames} Phoenix detector array after
 flat-fielding, correcting for the slit tilt, and background and
 nebular continuum removal. Intensity is given in units of the noise
 on a linear grey scale covering the full range of intensities. x-axis
 is wavelength in microns, y-axis is offset along the slit in
 arcsec. The y-axis increases towards the East. The horizontal arrows
 limit the optimal extraction aperture.}
\end{figure}

The resulting spectra for the three best nights are shown in
Fig.~\ref{fig:frames}, after subtraction of a small level of continuum
nebular emission. Wavelengths are given in air and in the observatory
rest frame.  There are at least four emission features observed near
the [Al\,{\sc vi}] wavelength given by \citet{cas00} of
3.659~$\mu$m. The three brightest features share simular
position-velocity structures.

Because the 1~arcsec uncertainty in acquisition is larger than the
slit width, spectra taken on different nights may not sample the same
region in the nebula, so we only coadded the frames taken with the
same instrumental setting (i.e. within the same night). Another reason
to avoid averaging all the spectra is to keep a good spectral
resolution: it can be inferred by inspection of Fig.~\ref{fig:frames}
that the night of July 30$^\mathrm{th}$ has the best line contrast,
even though we used the widest slit. The sharpest lines should be
obtained with the narrowest slit.  The quality of the spectrum from
July 30$^\mathrm{th}$ reflects that it was acquired under the best
weather conditions.

%There were no reference lines in the [Al\,{\sc vi}] frames\footnote{by
%frame we mean the detector image after exposure in the [Al\,{\sc vi}]
%settings} from the Th-Ar-Ne arc lamp with which to calibrate the
%dispersion law. 
%

%\footnote{by rows we mean detector lines in the dispersion direction}

No reference lines were visible in the calibration exposures taken
with a Th-Ar-Ne arc lamp in the [Al\,{\sc vi}] instrument
configuration. Instead we used emission lines from coadded sky spectra
extracted from the science observations (without differencing the
nodded frames).  We fit a straight line to the position of sky
emission features present in a model high-resolution sky spectrum
based on the HITRAN database \citep{rot92}. The accuracy of the
inferred dispersion law is checked {\em a-posteriori} by comparing
different nights and previous wavelength measurements. The raw spectra
are modulated by the atmospheric transmission (hereafter AT) spectrum,
which is reasonably smooth near [Al\,{\sc vi}] (AT does not show deep
troughs).  We correct for the AT modulation by dividing the object
frames by the spectrum of a standard star (HR~6789) grown along the
slit.

The optimal aperture for spectrum extraction in the spatial direction
along the slit was determined by varying the upper $y_\mathrm{up}$ and
lower $y_\mathrm{lo}$ rows of detector pixels.  We summed all the
signal in the detector within the rows $y_\mathrm{lo}$ and
$y_\mathrm{up}$, and estimate the noise level {\em a-posteriori}, from
the rms dispersion of the output spectrum in a region devoid of line
emission. A search in the 2-D parameter space $\{y_\mathrm{lo}$,
$y_\mathrm{up}\}$ for the best signal-to-noise spectrum gives the
optimal aperture indicated in Fig.~\ref{fig:frames}. We hereafter
refer to spectra extracted with this optimal aperture as collapsed
spectra.

\begin{table*}
 \caption{Observation log. S/N is the ratio of the peak specific intensity to the noise in the image.}
 \label{table:log}
 \begin{tabular}{lcccccc}
 \hline
  date  & slit  & S/N  &  integration & airmass & seeing$^1$ & weather$^1$  \\ 
  2003  & arcsec  &     &  s           &        &            &  \\ 
  \hline
May  7  &  0.25  & 14 & 84 $\times$ 180 &  $1.27 \rightarrow  1.03,  1.05 \rightarrow 1.4$ & 2 & 2  \\
Jul 26  &  0.17  & 12 & 40 $\times$ 300 &  $1.10\rightarrow 1.01\rightarrow 1.17$ & 3  & 2  \\
Jul 27  &  0.34  & 10 & 12 $\times$ 300 &  $1.45\rightarrow 2.08$ & 3  & 3 \\
Jul 30  &  0.34  & 12 & 12 $\times$ 300 &  $1.28\rightarrow 1.64$  & 1  & 1 \\
Jul 31  &  0.34  & 14 & 24 $\times$ 300 &  $1.08\rightarrow 1.01\rightarrow 1.02$  & 2 & 4 \\
  \hline
 \end{tabular}

\medskip

$^1$refers to a relative quality number, assigned by inspection, in which 1 is best.
\end{table*}

\section{Hyperfine spectral fits} \label{sec:hfsfits}

In Russell-Saunders coupling the magnetic field due to the nuclear
spin splits a given $\{L, J\}$ fine-structure level into hyperfine
levels, with the following energy shifts \citep{gla78},
\begin{eqnarray}
\lefteqn{ \Delta E(L,J,F,I) = \frac{h A_{L,J}}{2} K } \nonumber \\  & & +  h B_{L,J} \left[ K (K+1) - \frac{4}{3}I(I+1)J(J+1)\right],  
\end{eqnarray}
where 
\begin{equation}
K = F\,(F+1) - I\,(I+1) - J\,(J+1), 
\end{equation}
where $L$ stands for the electronic orbital angular momentum and $I$
is the nuclear spin. $h$ is the Planck constant.  $A_{L,J}$ and
$B_{L,J}$ are the magnetic-dipole and electric-quadrupole hyperfine
coupling constants, respectively. $F\,(F+1)$ is an eigenvalue of ${\bf
F}^2$, where ${\bf F}$ is the vectorial ${\bf I+J}$ operator.  The
relative intensities $S(\{J_1,F_1\},\{J_2,F_2\})$ of each hyperfine
component $\{I,J_1,F_1\} \leftarrow \{I,J_2,F_2\} $ can be derived
from
\begin{eqnarray}
\lefteqn{ S(\{I,J_1,F_1\},\{I,J_2,F_2\}) = } \nonumber \\ 
& &  (2F_1+1) (2F_2+1)   \left\{ \begin{array}{lll} F_2 & F_1  & 1 \\ J_1 &  J_2 &  I \end{array} \right\}^2,
\end{eqnarray}
with the  selection rule:
\begin{equation}
|F_1-F_2| \leq 1 \leq F_1 + F_2,
\end{equation}
where \{ \} is the six-$j$ symbol defined by \citet{bri94}.

The model hyperfine structure of [Al\,{\sc vi}] given in
Table~\ref{table:superstructure} derives from an ab-initio calculation
of the HFS coupling constants (carried out by one of us, PJS) using
the atomic structure code {\sc superstructure}
\citep[][]{eis74,cle97}:
\begin{equation}
\begin{array}{lr}
A^\mathrm{th}_{J=1}(^{27}\mathrm{Al}) = 0.2\,, &  \: A^\mathrm{th}_{J=2}(^{27}\mathrm{Al}) = 3461.8 \,, \\
 A^\mathrm{th}_{J=1}(^{26}\mathrm{Al}) = 0\,, &  \: A^\mathrm{th}_{J=2}(^{26}\mathrm{Al}) = 1333\,, \end{array}\label{eq:ath_26al}
\end{equation}
where all values are given in MHz. Table~\ref{table:superstructure}
does not include the electric-quadrupole terms in the hyperfine energy
shifts, because {\sc superstructure} does not currently predict the
electric-quadrupole hyperfine coupling constants.  The velocities in
Table~\ref{table:superstructure} are all given relative to the
strongest component of the stable isotope. The nuclear magnetic dipole
moments used in the calculation are $+3.64151~\mu_N$ for $^{27}$Al
\citep{rag89}, and $+2.804~\mu_N$ for $^{26}$Al \citep{coo96}, in
units of the nuclear magneton $\mu_N = e \hbar/ 2 m_p c$.

The hyperfine coupling constants given above were calculated in a
simple two configuration atomic model, 2s$^2$2p$^4$ and 2p$^6$.
Calculations were also made for more elaborate configuration bases and
with different orbital optimisation procedures leading to a range of
values for the hyperfine coupling constants. The results were all
within 100~MHz of the values quoted above but with no obvious
convergence to one particular best result.  We therefore quote the
results of the simplest calculation and adopt $\sigma^\mathrm{th} =
100$~MHz as the likely uncertainty in the theoretical result. This
uncertainty has to be compared to the difference
$|A^\mathrm{th}_{J=1}-A^\mathrm{th}_{J=2}|\sim 1000$, which is roughly
how these quantities enter the expression for the hyperfine energy
shifts. The theoretical $A^\mathrm{th} $ agree with observations
within $2 \sigma^\mathrm{th}$ (see below).  The high accuracy of the
calculation for $^{27}$Al should be carried over to $^{26}$Al since
they have the same electronic wave functions (to a very good
approximation). Therefore it is reasonable, within a $<10\%$
uncertainty on the HFS coupling constants, to use the theoretical
hyperfine constants for $^{26}$Al in a fitting procedure to look for
evidence of $^{26}$Al.

There is no allowance for isotopic mass shift neither in the ab-initio
calculations nor in the spectral fits. We assume both $^{26}$Al and
$^{27}$Al share the same line centroid.  The normal mass shift (NMS)
due to the difference in Rydberg constants between $^{26}$Al and
$^{27}$Al would cause the centroid of the 3.6$\mu$m line in $^{26}$Al
to be shifted to the red by 0.23 km s$^{-1}$ relative to
$^{27}$Al. The specific mass shift (SMS) is not known for [Al\,{\sc
vi}] but measurements have been made for the same transition in the
isoelectronic O\,{\sc i} \citep{den93}, which show that the ratio of
the total isotope shift to the normal mass shift ((NMS+SMS)/NMS) is
1.40 between $^{17}$O and $^{16}$O and 1.26 between $^{18}$O and
$^{17}$O. Adopting the larger of the these two values, we can estimate
that the total isotope shift of the $^{26}$Al centroid relative to
$^{27}$Al should be no more than 0.32~km~s$^{-1}$.

For both isotopes there are actually nine lines which in practice
reduce to five due to degeneracy in the hyperfine levels associated
with the $J=1$ state. This degeneracy is not exact but in practice the
lines lie within less than 0.1~km~s$^{-1}$ of each other, which is
much smaller than the typical linewidth of $\sigma \sim
8$~km~s$^{-1}$, so they can be taken to have the same velocity shift.
The number of components within each line is given in brackets in
Table~\ref{table:superstructure}.  We stress that
Table~\ref{table:superstructure} does not include the electric
quadrupole terms in the hyperfine energy shifts.

\begin{table}
 \caption{{\sc superstructure} ab-initio calculation of velocity
 shifts relative to the strongest component of $^{27}$Al, and
 corresponding relative intensities, for [$^{26}$Al\,{\sc vi}] and
[$^{27}$Al\,{\sc vi}].}
 \label{table:superstructure}
\begin{center}
 \begin{tabular}{lclc}
 \hline
\multicolumn{2}{c}{$^{27}$Al\,{\sc vi}} &  \multicolumn{2}{c}{$^{26}$Al\,{\sc vi} } \\
  rel. &     rel. &        rel. &      rel.\\
  vel. &    int. &      vel.  &     int. \\
  (km~s$^{-1}$)   &              &      (km~s$^{-1}$) &    \\
     0.0   &       5.0 (1)$^1$&    -14.8 &          9.0  (1)\\
   -58.0   &       4.0 (2)    &    -49.5 &          7.8  (2)\\
  -103.1  &       3.0 (3)    &    -79.3 &          6.6  (3)\\
  -135.2   &       2.0 (2)    &   -104.1 &          5.4  (2)\\
  -154.5   &       1.0 (1)    &   -107.3 &          4.2  (1)\\

  \hline
 \end{tabular}
\end{center}

\medskip

$^1$the number of sub-components blended together in each  velocity component.
\end{table}

We fit the [Al\,{\sc vi}] line profile $F_\lambda$ with the following
parametrised model,
\begin{eqnarray}
\lefteqn{ F_\lambda =  F_\circ+ \sum_{i=1}^{N_\mathrm{isotope}} 
\sum_{\mathrm{g}=1}^{N_\mathrm{gauss}} 
\sum_{F_1,F_2}  
S(\{I,J_1,F_1\},\{I,J_2,F_2\}) }  \nonumber \\ & & ~R_i ~R_g~  \exp \left(  -\frac{1}{2} \frac{(\lambda - \lambda_g(F_1,F_2))^2}{\sigma_g} \right),   
\end{eqnarray}
where $F_\circ$ is a constant baseline, $N_\mathrm{isotope}$ is the
number of isotopes (i.e. 1 or 2 in this case), $N_\mathrm{gauss}$ is
the number of Gaussians used to represent the fit (1 or 2), $R_i$ is
an overall amplitude for isotope $i$, $R_g$ is the relative amplitude
of additional Gaussians relative to the first. Thus the first Gaussian
component for isotope $i$ has $R_{g=1} = 1$, and $R_i$ is constant for
all $g$ components of isotope $i$.  $\lambda_g(F_1,F_2)$ is the
Gaussian centroid of each hyperfine component,
\begin{eqnarray} \label{eq:model2}
\lefteqn{  \lambda_g(F_1,F_2)  = } \\ & &  \lambda_\circ + \Delta \lambda_g   + \frac{c~ h}{\Delta E(L,J_2,F_2,I) - \Delta E(L,J_1,F_1,I)},
\end{eqnarray}
where $\lambda_\circ$ is a reference wavelength which does not
necessarily match the fine-structure transition, which instead
corresponds to the overall centroid of the line, i.e. the average of
each HFS component weighted by its flux. $\Delta \lambda_g$ is an
offset to describe the velocity profile with Gaussian $g$ (one
Gaussian has $\Delta \lambda = 0$).

The optimisation was carried-out by minimising $\chi^2 = \sum_j
(F(\lambda_j)-F_m(\lambda_j))^2/\sigma^2_F$ in two steps. We perform
an initial heuristic search of the global minimum with the {\tt
pikaia} genetic algorithm of \citet{cha95}, and then optimise with the
variable-metric routine {\tt MIGRAD} of the {\tt Minuit} package from
\citet{CERN}. We took the precaution of cross-checking the {\tt
MIGRAD} results with the downhill simplex method {\tt amoeba}
\citep{pre86}, which we observe to be much slower and far less robust
than {\tt MIGRAD} ({\tt amoeba} requires fine tuning of the input
simplex and tolerance parameters).  Errors on individual parameters
are estimated by searching parameter space for the $\Delta~\chi^2=1$
contour.

\begin{figure}
\includegraphics[width=8.5cm,height=!]{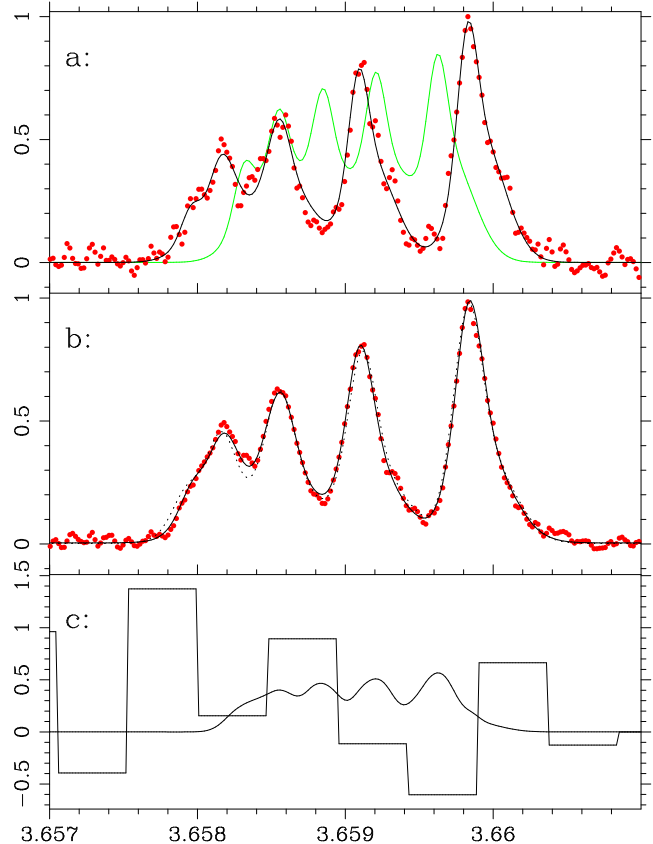}
 \caption{\label{fig:spectrum} a) Points: collapsed spectrum of
 [Al\,{\sc vi}] from 30-06-2003, with the optimal extraction
 aperture. Solid line: the best fit with two Gaussians per component,
 and the parameters given in Table~\ref{table:fits}, without
 contribution from $^{26}$Al.  Grey solid line: the profile of
 $^{26}$Al, had it been present at a level giving an isotope ratio of
 1. b) Points: coadded spectrum. Solid-line: combined
 model. Dotted-line: combined model without electric-quadrupole
 hyperfine splitting. c) Histogram: binned residuals, excluding the
 $^{26}$Al fits. Solid line: combined $^{26}$Al fit.}
\end{figure}

The resulting observed spectrum and model line profile are shown on
Fig.~\ref{fig:spectrum}. We show the case of the optimal-extraction
spectrum from July 30$^\mathrm{th}$ in Fig.~\ref{fig:spectrum}a,
together with an indication of the hyperfine splitting of $^{26}$Al,
had it been present. The coadded spectrum is compared to the fits in
Fig.~\ref{fig:spectrum}b, where it can be appreciated that the
inclusion of the electric quadrupole hyperfine terms improves the
fit. It can be verified by inspection that the solid line, with $B$
terms, is appreciably closer to the data than the dotted line, without
$B$ terms. The residuals are shown on Fig.~\ref{fig:spectrum}c, with
the formal $^{26}$Al fit.

Our best-fit line profiles are summarised in
Table~\ref{table:fits}. The $^{26}$Al hyperfine coupling constants
were kept fixed at their theoretical values, as specified in
Eq.~\ref{eq:ath_26al}. We list reduced $\chi^2$ as an indicator of
goodness of fit: values much less than 1 reveal that we are fitting
the noise with an excessive number of free parameters. We nonetheless
include these fits in the list with the goal of combining the results
from all nights.

%But this constraint was relaxed for the {\tt minuit} systematic
%optimization.  Thus the values we report for $A_J(^26\mathrm{Al})$
%correspond at least to a local minimum in the fit. The lack of a
%secure $^26$Al detection precludes giving confidence in its HFS
%coupling constants.

We can use the information that the HFS coupling constants are the
same on each night to perform a second run of the fitting procedure,
and fix the HFS constants to the average given in
Table~\ref{table:fits}. This allows improved estimates of the fine
structure centroid, as well as tighter limits on the abundance of
$^{26}$Al relative to $^{27}$Al.  The results of this second run of
fits are summarised in Table~\ref{table:iso}. The average value for
the fine structure centroid includes a correction for the heliocentric
systemic velocity of NGC~6302 of $-$35.0~km~s$^{-1}$ \citep{cas00},
and is accurate to within 1~km~s$^{-1}$.  The constraints we can
place on the Al isotope ratio are summarised under columns
$R_\mathrm{iso}^a$ and $R_\mathrm{iso}^b$.  The 1~$\sigma$ values and
upper limits indicated under $R_\mathrm{iso}^a$ are formal indicators
of the relative limits set by each spectra, and are derived from the
first fitting procedure, with free HFS constants.  In this case we
cannot use the error estimates derived from the $\Delta \chi^2=1$
contour because the positivity requisite on $R_\mathrm{iso}$ precludes
reaching the global $\chi^2$ minimum with certainty. But we relaxed
the positivity constraint in the second fit, fixing the HFS
constants. The results are listed in Table~\ref{table:iso}, under the
column $R_\mathrm{iso}^b$.  The weighted average for $100\times
R_\mathrm{iso}$ is $0.6 \pm 0.8$.

We estimate an upper limit on $R_\mathrm{iso}$ by generating a
synthetic spectrum with $R_\mathrm{iso} = 0.03$, and repeating the
fitting procedure, for 100 different realisations of Gaussian noise,
at the same level as that of the collapsed spectrum for 30-06-2003
(which has the best S/N). This Monte-Carlo error analysis shows we can
recover the input isotope ratio at 2-$\sigma$: $100\times
R_\mathrm{iso} = 3.1 \pm 1.6$. Another simulation with the noise level
of the residual spectrum shown on Fig.~\ref{fig:spectrum} gives
$100\times R_\mathrm{iso} = 3.0 \pm 0.8$. These simulations and the
combined measurement of $R_\mathrm{iso}$ from Table~\ref{table:iso}
are in agreement, which allows us to place the following 3-$\sigma$
upper limit:
\begin{equation}
R_\mathrm{iso} < \langle R_\mathrm{iso} \rangle + 3\sigma = 3.0~10^{-2}.
\end{equation}

%The $^{26}$Al hyperfine coupling constants $A_J(^26\mathrm{Al})$ were
%kept fixed at their theoretical value for the {\tt pikaia} heuristic
%global optimisation, but this constraint was relaxed for the {\tt
%minuit} systematic optimisation. Thus the values we report for
%$A_J(^26\mathrm{Al})$ correspond at least to a local minimum in the
%fit. The lack of a secure $^26$Al detection precludes giving
%confidence in its HFS coupling constants.

\begin{figure}
\begin{center}
\includegraphics[width=7cm,height=!]{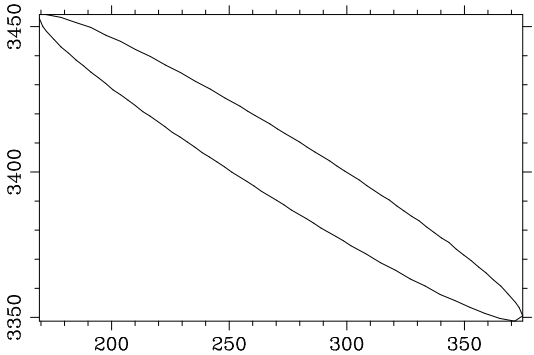}
\end{center}
 \caption{\label{fig:bias} Bias due to the correlation of the
 free-parameters for the HFS constants $A_{J=2}(^{27}\mathrm{Al})$
 (y-axis) and $A_{J=1} (^{27}\mathrm{Al})$ (x-axis), from the
 collapsed spectrum of July~30th.}
\end{figure}

The uncertainty in the measured HFS coupling constants, relative to
the optimal value, is rather large compared to that of the central
wavelengths. This is due to a significant statistical bias in the
values of $A_{J=2}(^{27}\mathrm{Al})$ and $A_{J=1}
(^{27}\mathrm{Al})$. Fig.~\ref{fig:bias} is a 2-D slice in parameter
space showing the correlation of both constants. With the neglect of
the quadrupole HFS constants $B_{L,J}$, the bias is much stronger and
the uncertainty on the magnetic dipole constants is much larger ($\sim
5$ times larger): $A_{J=1}(^{27}\mathrm{Al})$ = $149^{+171}_{-444}$,
$A_{J=2}(^{27}\mathrm{Al})$ = $3499^{+224}_{-87}$.  In the absence of
the electric quadrupole terms, the hyperfine energy shifts depend on
the $A_{L,J}$ constants approximately through their difference,
$A_{J=2}(^{27}\mathrm{Al}) - A_{J=1} (^{27}\mathrm{Al})$.
Notwithstanding this difficulty, the observed constants are close to
the theoretical values used to produce
Table~\ref{table:superstructure}.

The reasons why we are confident on our detection of the
electric-quadrupole hyperfine splitting are as follows.
\begin{enumerate}
\item The fit to the line profile significantly improves, with reduced
$\chi^2$ increasing by more than 0.1 without the
$B_J(^{27}\mathrm{Al})$ constants, systematically for all nights. For
example, in the case of the collapsed spectrum for
July~30$^\mathrm{th}$, $\chi^2/\nu$ rises from 1.08 to 1.20.
\item The improvement in the fit is visible to the eye by inspection
of Fig.~\ref{fig:spectrum}, where it can be appreciated that the fit without 
$B_J(^{27}\mathrm{Al})$ gives an excess in the blue. 
\item The two electric quadrupole constants are measured with
accuracies of $6\sigma$ and $10\sigma$. 
\item The average isotope ratio without the quadrupole terms is
  $(-1.41 \pm 0.59)\times 10^{-2}$, which reflects a tendency to
  compensate for the misfit with a negative, and spurious, amplitude
  for the rare isotope.
\end{enumerate}

The intrinsic profile of [Al\,{\sc vi}] is manifestly very narrow. A
single Gaussian fit to the spectrum from 30-06-2003, with a 6-row
extraction centred on the peak of emission along the slit, gives a
width of $\sigma = 7.31\pm0.24$~km~s$^{-1}$, or a FWHM of
17.21$\pm$0.56~km~s$^{-1}$. This spectrum was acquired with the widest
slit, and a resolving power $R=40\,000$. We cannot give a precise
measure of the instrumental resolution because of the lack of arc
lines near [Al\,{\sc vi}]. But assuming $3~10^5/R$~km~s$^{-1}$
corresponds to the FWHM instrumental resolution within 20\%, we can
give an estimate of the de-convolved line width of
$15.5\pm1.0$~km~s$^{-1}$ FWHM. A comparison with emission lines from
lighter species is deferred to a forthcoming article.

%If entirely due to thermal broadening, the linewidth of [Al\,{\sc vi}]
%gives a temperature of $140\,000\pm18\,000$~K.

%\begin{table*}
% \small
% \caption{HFS fits to the observations}
% \label{table:fits}
% \begin{tabular}{lllllllllll}
% \hline
%                       &       &             \multicolumn{3}{c}{$^{27}\mathrm{Al}$}                                        & \multicolumn{3}{c}{$^{26}\mathrm{Al}$}   &    &  &      isotope          \\
% date, 2003             & noise$^1$ &  $\lambda_\circ$    & $A_{L,J=1}$  &  $A_{L,J=2}$   &  $\lambda_\circ$ & $A_{L,J=1}$  &  $A_{L,J=2}$   &    $\sigma_1$ & $R_2, \sigma_2$ &  ratio$\times$100   \\
%30 July, peak$^{2}$     &  0.0241 &  $3.6590083(^{+16}_{-17})$ & $396^{+53}_{-61}$   &  $3375^{+32}_{-28}$ &   $0^{+188}_{-190}$ &   $3.659706(^{+18}_{-17})$                 &   $1262^{+107}_{-103}$  & $6.27^{+0.24}_{-0.26}$ &   0.32,$13.65^{+1.25}_{-1.03}$ &  $3.34^{+0.89}_{-0.91}$ \\
%  \hline
% \end{tabular}
%
%\medskip
%
%$^1$noise used to assess the significance of the fits, normalised to the peak flux density (i.e. S/N = 1/noise). 
%$^2$1-row spectrum at the peak of emission.
%$^3$collapsed slit (optimal S/N). 
%\end{table*}

\begin{table*}
 \caption{HFS fits to the observations}
 \label{table:fits}
 \begin{center} \small
 \begin{tabular}{llllllllll}
 \hline
                       &       &         &              \multicolumn{5}{c}{$^{27}\mathrm{Al}$}                                      &             &           \\
 date,               & noise$^a$ & $\chi^2/\nu$ &   $\lambda_\circ^d$    & $A_{L,J=1}^e$  &  $A_{L,J=2}$     &  $B_{L,J=1}$  &  $B_{L,J=2}$     &    {\sc fwhm}$_1$,$\Delta \lambda_1$  & $R_2$, {\sc fwhm}$_2$  \medskip   \\
2003                   &      &                &   $\mu$m     & MHz   &  MHz  &  MHz  &  MHz     &     km~s$^{-1},10^{-5}~\mu$m  &    $-,$km~s$^{-1}$  \medskip   \\
7 May$^{c}$ &   $ 4.4~10^{-2} $  &  1.02      & $ 3.658715^{+35}_{-22}$      & $     -0.1^{+162}_{-153}$      & $   3489.1^{+79}_{-82}$      & $     -0.0^{+13}_{-13}$      & $      6.9^{+2}_{-2}$     & $  18.3^{+  1.0}_{-  0.9},-15.1 $   & $ 0.56, 36.7^{+  2.0}_{-  2.7}$    \medskip\\    
26 July$^{c}$  &  $ 4.5~10^{-2} $  &  0.98      & $ 3.659112^{+27}_{-27}$      & $      0.0^{+243}_{-231}$      & $   3499.1^{+117}_{-122}$      & $      0.0^{+22}_{-21}$      & $     10.5^{+4}_{-4}$     & $  23.9^{+  0.8}_{-  0.9},-5.4 $   & $ 0.20, 65.2^{+ 11.6}_{- 10.7}$  \medskip\\    
27 July$^{c}$  &  $ 7.0~10^{-2} $  &  0.87      & $ 3.659341^{+ 9}_{-11}$      & $    417.7^{+123}_{-121}$      & $   3332.9^{+65}_{-66}$      & $    -54.6^{+11}_{-10}$      & $     16.9^{+2}_{-2}$     & $  22.3^{+  0.7}_{-  0.7},-30.3 $   & $ 0.26, 11.6^{+  2.2}_{-  1.6}$   \medskip \\    
30 July$^{b}$  &  $ 6.3~10^{-2} $  &  0.80      & $ 3.659247^{+16}_{-23}$      & $    258.2^{+150}_{-249}$      & $   3390.2^{+124}_{-76}$      & $    -19.7^{+15}_{-18}$      & $      8.7^{+3}_{-3}$     & $  15.9^{+  0.7}_{-  0.8},-18.8 $   & $ 0.13,  8.8^{+  3.8}_{-  2.5}$ \medskip\\    
30 July$^{c}$  &   $ 4.1~10^{-2} $  &  1.08      & $ 3.659125^{+13}_{-11}$      & $    276.9^{+98}_{-108}$      & $   3399.2^{+55}_{-50}$      & $    -29.8^{+8}_{-9}$      & $      9.3^{+2}_{-2}$     & $  13.1^{+  0.8}_{-  0.8},-9.9 $   & $ 0.82, 31.4^{+  0.9}_{-  0.8}$   \medskip\\   
31 July$^{c}$  &   $ 5.3~10^{-2} $  &  0.68      & $ 3.659365^{+32}_{-58}$      & $   -117.0^{+446}_{-238}$      & $   3575.9^{+122}_{-224}$      & $    -17.2^{+50}_{-17}$      & $      8.8^{+3}_{-9}$     & $  17.1^{+  0.6}_{-  0.7},-30.2 $   & $ 0.11, 26.7^{+  8.9}_{-  6.1}$ \medskip  \\    
     \multicolumn{4}{c}{average}     &   $ 235.5 \pm 62.8$  &   $3410.1 \pm 32.4$  &   $-28.0 \pm   5.3$   &   $10.7 \pm  1.0 $  &  &  \medskip\\
  \hline
 \end{tabular}
 \end{center}

\medskip
\raggedright
$^a$noise used to assess the significance of the fits, normalised to the peak flux density (i.e. S/N = 1/noise). \\
$^b$6-row spectrum centred on the  peak of emission.\\
$^c$collapsed slit (optimal S/N). \\
$^d$Uncertainties on $\lambda_\circ$   refer to the last decimal places. $\lambda_\circ$   does not match  the fine-structure  centroid (see text and Eq.~\ref{eq:model2}). \\
$^e$Uncertainties on all quantities refer to the usual 68.3~\%
confidence interval (i.e., 1~$\sigma$ for 1 parameter).  
%$^5$this spectrum is affected by inaccurate sky line subtraction due to poor weather \\ 
\end{table*}

\begin{table}
 \caption{The fine structure centroid and limits on the Al isotope ratio.}
 \label{table:iso}
 \begin{center} \small
 \begin{tabular}{llll}
 \hline
 date, 2003      &    $100\times R_\mathrm{iso}^1$ &  $100\times R_\mathrm{iso}^4$ & $\lambda_\mathrm{FS}^5$ \\
7 May$^{3}$      &   1.5$\pm$2.5  &    4.8$\pm$ 2.0    & 3.659273(35)    \\    
26 July$^{3}$    &   $ <   0.67$  &   -0.4$\pm$ 2.0    & 3.659427(19)  \\    
27 July$^{3}$    &   1.8$\pm$2.6  &   -1.5$\pm$ 2.3    & 3.659415(22)   \\               
30 July$^{2}$    &   1.6$\pm$2.5  &    2.9$\pm$ 2.3    & 3.659392(22)         \\    
30 July$^{3}$     &   $  <  0.88$  &   -0.9$\pm$ 1.7    &  3.659410(10)   \\          
31 July$^{3}$    &   $ < 0.72 $   &   -1.2$\pm$ 2.2    & 3.659427(40)            \\      
   average       &      --        &    0.6$\pm$ 0.8    &  3.659405(7)  \\
  \hline
 \end{tabular}
 \end{center}

\medskip
\raggedright
$^1$formal 1-$\sigma$ upper limit (see text for accurate upper limits). \\
$^2,$ and $^3$ same as Table~\ref{table:fits}\\
$^4$best fit isotopic ratio, using fixed HFS constants  \\
$^5$rest wavelength in $\mu$m and in air.  
\end{table}

%FS centroid  #vobj = -39
%[  3.659321   3.659476   3.659463   3.659441   3.659458   3.659476]
%[3.5e-05 1.9e-05 2.2e-05 2.2e-05 1e-05 4e-05]
%  3.65945388311682     7.38341413080541e-06   

%FS centroid  #vobj = -35
%[  3.659273   3.659427   3.659415   3.659392 3.65941   3.659427]
%[3.5e-05 1.9e-05 2.2e-05 2.2e-05 1e-05 4e-05]
%  3.65940558540063     7.38341413080541e-06   

%[3.5e-05 1.9e-05 2.2e-05 2.2e-05 1e-05 4e-05]
%  3.65940558540063     7.38341413080541e-06   

%  NGC6302_AlVI_07_05_2003_spec_collapse.dat &       1.1 \pm      1.7    & 3.658645 (4)   \\    
%  NGC6302_AlVI_26_06_2003_spec_collapse.dat &      -1.6 \pm      1.7    & 3.659085 (5)   \\    
%  NGC6302_AlVI_27_06_2003_spec_collapse.dat &      -4.9 \pm      2.2    & 3.659084 (12)   \\    
%  NGC6302_AlVI_30_06_2003_spec_rows1.dat &         -6.4 \pm      2.6    & 3.659065 (4)   \\    
%  NGC6302_AlVI_31_06_2003_spec_collapse.dat &      -7.1 \pm      2.0    & 3.659119 (2)   \\    

\section{Conclusions} \label{sec:conc}

We have identified the multiple components near [Al\,{\sc vi}] as due
to the HFS splitting of $^{27}$Al. Theory agrees with the observed
magnetic dipole HFS coupling constants within the uncertainties,
giving support for the use of theoretical constants in the modelling
of ionic lines profiles. 

We provide the first measurements of electric quadrupole hyperfine
coupling constants for any atomic transition in any astrophysical
object. We discuss the spectroscopic importance of the quadrupole
terms. The inclusion of the quadrupole terms improves the measurement
of the magnetic dipole constants, which are otherwised affected by a
statistical bias.

As an application we have set a 3$\sigma$ upper limit on the aluminum
isotopic ratio, $^{26}$Al/$^{27}$Al $< 1/33$.  This is the most
stringent upper limit on the relative $^{26}$Al abundance in any
astrophysical object to date.

However, the accuracy of our measurement is short of quantifying
$^{26}$Al production in AGB stars. The expected isotopic ratio at the
tip of the AGB is at most 1/37, from the ratio of the $^{26}$Al and
$^{27}$Al yields in the 6~M$_{\odot}$ models of \citet{for97}. The
progenitor mass of NGC~6302 is about 5--6~M$_{\odot}$ from the data
summarised in \citet{cas00}. But the predicted $^{26}$Al/$^{27}$Al
ratio lies below our $3~\sigma$ upper limit. We can only discard
$R_\mathrm{iso} = 1/37$ at 2.5~$\sigma$.  Doubling our integration on
[Al\,{\sc vi}] in NGC~6302 would allow a firm test on the theoretical
predictions.

To establish useful constraints on the $^{26}$Al production by AGB
stars we must deepen our observations of NGC6302, and extend the
analysis to other targets. Only PNe and symbiotic stars have moderate
expansion velocities and photoionised coronal-line regions, offering
narrow emission line profiles in high-excitation species, which are
otherwise thermally broadened in the Sun.

\section*{Acknowledgments}

Many thanks to the Gemini support team, in particular Tom Geballe, Bob
Blum and Bernadette Rodgers. Thanks also to Andr\'es Jordan for the
{\tt PDL::Minuit} package. S.C acknowledges support from Fondecyt
grant 1030805, and from the Chilean Center for Astrophysics FONDAP
15010003. Based on observations obtained at the Gemini Observatory,
which is operated by the Association of Universities for Research in
Astronomy, Inc., under a cooperative agreement with the NSF on behalf
of the Gemini partnership: the National Science Foundation (United
States), the Particle Physics and Astronomy Research Council (United
Kingdom), the National Research Council (Canada), CONICYT (Chile), the
Australian Research Council (Australia), CNPq (Brazil) and CONICET
(Argentina).

\bsp

\label{lastpage}

\end{document}